\documentclass[onecolumn, 10pt, superscriptaddress, notitlepage, nofootinbib, longbibliography]{revtex4-1}
\usepackage[utf8]{inputenc}
\usepackage{amsfonts,amsmath,amssymb}
\usepackage{graphicx}
\usepackage{soul}
\usepackage{color}
\usepackage{placeins}
\usepackage[colorlinks=true, breaklinks=true, linkcolor=darkblue, citecolor=darkblue, urlcolor=darkblue]{hyperref}
\definecolor{darkblue}{rgb}{0, 0, 0.8}

\begin{document}

\title{Nonlinear optics in the fractional quantum Hall regime}

\author{Patrick Kn\"uppel*}
\affiliation{Institute of Quantum Electroncis, ETH Z\"urich,\\
CH-8093 Z\"urich, Switzerland}

\author{Sylvain Ravets*}
\affiliation{Institute of Quantum Electroncis, ETH Z\"urich,\\
CH-8093 Z\"urich, Switzerland}
\affiliation{CNRS Centre for Nanoscience and Nanotechnology, \\
Universit\'e Paris-Sud, Universit\'e Paris-Saclay, 91120 Palaiseau, France}

\author{Martin Kroner}
\affiliation{Institute of Quantum Electroncis, ETH Z\"urich,\\
CH-8093 Z\"urich, Switzerland}

\author{Stefan F\"alt}
\affiliation{Institute of Quantum Electroncis, ETH Z\"urich,\\
CH-8093 Z\"urich, Switzerland}
\affiliation{Solid State Physics Laboratory, ETH Z\"urich,\\
CH-8093 Z\"urich, Switzerland\\
*These authors contributed equally to this work.}

\author{Werner Wegscheider}
\affiliation{Solid State Physics Laboratory, ETH Z\"urich,\\
CH-8093 Z\"urich, Switzerland\\
*These authors contributed equally to this work.}

\author{Atac Imamoglu}
\affiliation{Institute of Quantum Electroncis, ETH Z\"urich,\\
CH-8093 Z\"urich, Switzerland}

\maketitle

{\bf
Engineering strong interactions between optical photons is a great challenge for quantum science. Envisioned applications range from the realization of photonic gates for quantum information processing \cite{obrien_photonic_2009} to synthesis of photonic quantum materials for investigation of strongly-correlated driven-dissipative systems \cite{carusotto_quantum_2013}. Polaritonics, based on the strong coupling of photons to atomic or electronic excitations in an optical resonator, has emerged as a promising approach to implement those tasks \cite{sanvitto_road_2016}. Recent experiments demonstrated the onset of quantum correlations in the exciton-polariton system \cite{munoz-matutano_emergence_2019, delteil_towards_2019}, showing that strong polariton blockade \cite{verger_polariton_2006} could be achieved if interactions were an order of magnitude stronger. Here, we report time resolved four-wave mixing experiments on a two-dimensional electron system embedded in an optical cavity \cite{smolka_cavity_2014}, demonstrating that polariton-polariton interactions are strongly enhanced when the electrons are initially in a fractional quantum Hall state. Our experiments indicate that in addition to strong correlations in the electronic ground state, exciton-electron interactions leading to the formation of polaron polaritons \cite{schmidt_fermi_2012, sidler_fermi_2017, efimkin_exciton-polarons_2018, ravets_polaron_2018} play a key role in enhancing the nonlinear optical response. Besides potential applications in realization of strongly interacting photonic systems, our findings suggest that nonlinear optical measurements could provide information about fractional quantum Hall states that is not accessible in linear optical response.
}

Polaritons have recently attracted considerable interest, motivated by the fact that their interactions can be engineered almost at will through the tunability of their matter component. For example, strongly interacting Rydberg polaritons have recently been obtained using the nonlinear behavior of Rydberg excitations in an ensemble of atoms \cite{peyronel_quantum_2012}, which led to the demonstration of Rydberg polariton blockade \cite{jia_strongly_2018} where the presence of a single polariton in a well-delimited region of space prevents the resonant injection of other polaritons. In parallel, efforts are being made to realize polariton blockade in condensed matter systems that hold great potential for realizing compact and integrated synthetic quantum materials \cite{sanvitto_road_2016}. Exciton polaritons in semiconductor materials are part light part matter particles that arise from the strong coupling of a quantum well exciton and a cavity photon \cite{deng_exciton-polariton_2010}. These photonic particles inherit a nonlinear behavior from exciton-exciton interactions \cite{carusotto_quantum_2013, ciuti_role_1998, tassone_exciton-exciton_1999}. For efficient blockade to be obtained, the nonlinearity $U$ needs to be greater than the inverse lifetime $\gamma$ of the polaritons \cite{verger_polariton_2006}. Recent state-of-the art experiments based on photon correlation measurements in semi-integrated microcavities attained optimized values of the ratio $U/\gamma \simeq 0.1$ in a photonic dot with about $3~{\rm \mu m}^2$ area \cite{munoz-matutano_emergence_2019,delteil_towards_2019}. These experiments represent the culmination of decade long technological developments aimed at increasing $U/\gamma$ through reducing the photonic mode area \cite{el_daif_polariton_2006, ferrier_interactions_2011, besga_polariton_2015} as well as increasing the lifetime \cite{sun_bose-einstein_2017}. Recently, several possibilities have been explored for enhancing $U$ through an increase of exciton-exciton interactions, focusing either on biexciton Feshbach resonance \cite{carusotto_feshbach_2010, takemura_polaritonic_2014} or on excitons with a permanent dipole moment \cite{cristofolini_coupling_2012, byrnes_effective_2014, nalitov_voltage_2014, rosenberg_strongly_2018, togan_enhanced_2018}. The experiments we report here reveal a hitherto unexplored mechanism for optical nonlinearity emerging for polaritonic excitations out of a two dimensional electron system (2DES) in the fractional quantum Hall (FQHE) regime. More specifically, using time resolved four-wave mixing (FWM) experiments \cite{boyd_nonlinear_2008, smallwood_multidimensional_2018}, we find that polariton-polariton interactions $U$ can be enhanced by more than an order of magnitude around the fractional state at filling factor $\nu=2/5$ as compared to other neighboring compressible states. The interplay between photonic excitations and a 2DES is an exciting field \cite{kukushkin_spin_1999, byszewski_optical_2006, groshaus_absorption_2007, bar-joseph_trions_2005, hayakawa_real-space_2013} with open problems, among others, concerning the relation between transport and optics \cite{bartolo_vacuum-dressed_2018, paravicini-bagliani_magneto-transport_2019, cotlet_transport_2018} and the description of exciton-electron interactions in a magnetic field \cite{efimkin_exciton-polarons_2018}.

We study a semiconductor heterostructure that features, at the center of an optical microcavity, a GaAs QW containing an electron system of density $n_e = 3 \times 10^{10}\rm{cm}^{-2}$ (see Methods). In the presence of a 2DES, electron-exciton interactions modify the excitation spectrum and pioneering studies showed the existence of the strong coupling regime \cite{rapaport_negatively_2000, rapaport_negatively_2001}. A consistent description of the new excitonic excitations that emerge in the presence of a 2DES was first provided by R.~Suris \cite{suris_correlation_2003}, and were later termed exciton-polarons \cite{schmidt_fermi_2012, sidler_fermi_2017, efimkin_exciton-polarons_2018}. Strong coupling to an optical mode in turn results in the formation of polaron-polaritons as the elementary excitations of the 2DES-cavity system \cite{sidler_fermi_2017,ravets_polaron_2018}. Under an external magnetic field $B$ orthogonal to the 2DES surface, discrete Landau levels ${\rm LL}n_e$ (${\rm LL}n_{\rm hh}$) form out of the conduction (heavy-hole valence) band. We focus, in this article, on resonant optical excitations from the lowest Landau level ${\rm LL}0$. As we increase $B$, the filling factor $\nu$ of ${\rm LL}0$ decreases, allowing us to reach the fractional quantum Hall regime. We address transitions to the $\left | \uparrow \right >$ ($\left | \downarrow \right >$) ${\rm LL}0$ subband using $\sigma^-$ ($\sigma^+$) circularly-polarized light.

\begin{figure}
\centering
\includegraphics[width=120mm]{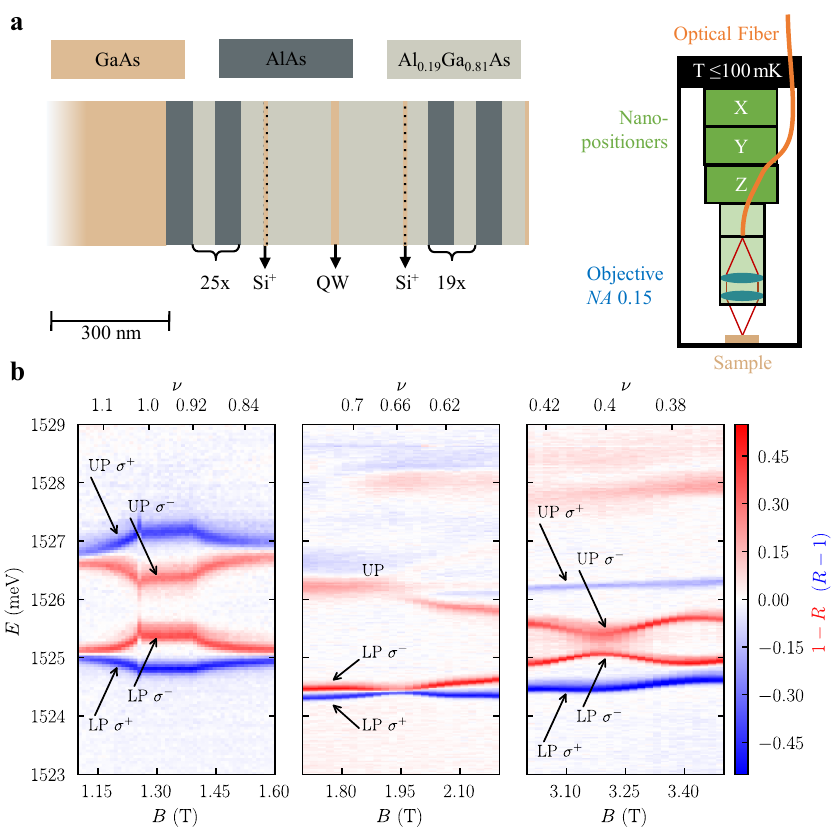}
\caption{{\bf Quantum Hall polaritons.}
{\bf a},~Sample structure and experimental setup for magneto-optical measurements at~mK temperatures. {\bf b},~White light reflectivity spectra recorded around filling factors $\nu=1$, $2/3$ and $2/5$. The plots show the difference between two spectra obtained separately using $\sigma^-$ and $\sigma^+$ polarized light.}
\label{fig:Fig1}
\end{figure}

We first characterize our sample using optical spectroscopy in the low-power (linear) regime. The sample is mounted inside a dilution refrigerator with fibered optical access, as shown in Fig.~\ref{fig:Fig1}a. We record reflectivity spectra for several values of $\nu$ using circularly-polarized light from a broadband light source. Fig.~\ref{fig:Fig1}b plots an overview of the polaron-polariton lines for our system, obtained by calculating the difference between the spectra measured using $\sigma^-$ (red) and $\sigma^+$ (blue) polarized light. We observe generic strong dispersion of the polariton energies with magnetic field around integer and fractional values of $\nu$. This striking behavior of the linear optical spectrum stems from strong modification of electron-exciton interactions in and around gapped quantum Hall states~\cite{ravets_polaron_2018}, which in turn leads to a $\nu$-dependent modification of the cavity-polaron coupling strength. Especially striking are the energy shifts experienced by the lower polariton in $\sigma^-$ polarization (${\rm LP}_{\sigma^-}$) at filling factors $\nu=1$, $2/3$ and $2/5$. Since the optical spectrum strongly depends on $\nu$, we refer to optical excitations in this system as quantum Hall polaritons.

\begin{figure}
\centering
\includegraphics[width=120mm]{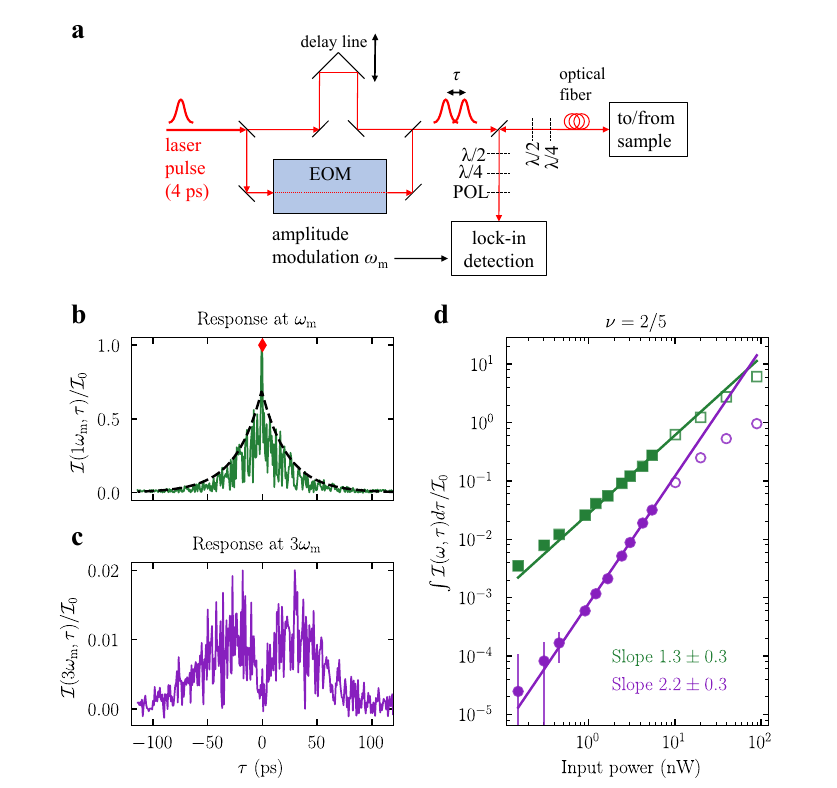}
\caption{{\bf Time resolved measurement of interactions between polaron polaritons.} {\bf a},~Experimental setup. Two laser pulses separated by a variable delay $\tau$ generate an induced polarization in the sample. The emitted photons are sent onto an APD where linear contributions are separated from nonlinear contributions using an electro-optical modulator (EOM) to modulate one of the beams in amplitude. {\bf b},~Typical linear and {\bf c},~nonlinear interference signals obtained for $\nu=2/5$ $(B=3.145\,{\rm T})$. All data is normalized to the maximal value of the linear response (red diamond). The dashed black line in {\bf b} shows a double-sided exponential decay fitted to the envelope of the linear response to obtain $T_\mathrm{LP}$. The input average power was set to $I_2=2\,\mathrm{nW}$. {\bf d},~Evolution of the linear (green squares) and nonlinear (purple circles) signal integral values as a function of the incident optical power (double logarithmic plot). The input power is given as the average power of the delayed pulse (i.e. $I_2$), and the errorbars correspond to the statistical error on the counts only. We fit the data before saturation of the nonlinearity (full circles) by a power law (green and purple lines). Errors on the power law exponents are dominated by systematic errors on the input power.}
\label{fig:Fig2}
\end{figure}

Figure~\ref{fig:Fig2}a shows the principle of the time-resolved experiment we use to characterize our sample in the nonlinear regime. We use a pulsed Ti:Sapphire laser with a $T_{\rm pulse} = 4\,{\rm ps}$ pulse duration, a 76\,MHz repetition rate and center frequency tuned to the ${\rm LP}_{\sigma^-}$ resonance. We split the laser into two paths and introduce a variable time delay $\tau$ between the two pulses. For optical excitation, we recombine both beams onto a beam splitter and couple the linearly-polarized light into an optical fiber routed to the sample. The excitation light is then focused onto the sample surface using a low NA objective. The total field incident on the sample is given by $E(t,\tau)=E_1(t)+E_2(t,\tau)$, where the average intensities of the two beams are chosen to be equal. For detection, we collect the generated resonance fluorescence using the same fiber as the one used for excitation, and we filter out the laser background light by detecting along the cross-polarized axis. The collected light is finally sent onto an avalanche photodiode (APD) for detection. Modeling our system as a third order nonlinear medium, we can expand the total intensity reaching the photodetector $I_{\rm det}$ as:
\begin{equation}
\label{Eq:Itot}
\begin{aligned}
I_{\rm det}(t,\tau) \propto \epsilon_0 \left | P^{(1)}_1(t) + P^{(1)}_2(t, \tau) + P^{(3)}(t, \tau) \right| ^2, \
\end{aligned}
\end{equation}
where the linear $P^{(1)}_{1,2}(t)$ (nonlinear $P^{(3)}(t)$) polarizations are the inverse Fourier transforms of $P^{(1)}_{1,2}(\omega)$ ($P^{(3)}(\omega)$) \cite{boyd_nonlinear_2008}. To isolate weaker nonlinear terms $\propto {P^{(1)*}_{i}} P^{(3)}$ from the dominant linear contributions $\propto {P^{(1)*}_{i}} {P^{(1)}_{j}}$ $(i,j=1,2)$, we modulate the field amplitude $E_1(t)$ sinusoidally at frequency $\omega_{\rm m}$. By calculating $\mathcal{I}(\omega, \tau)$, the Fourier transform of $I_{\rm det}(t,\tau)$, we can separate different terms: the (mostly) linear term $\mathcal{I}(\omega_{\rm m}, \tau)$ and the nonlinear term $\mathcal{I}(3 \omega_{\rm m}, \tau)$ (see Methods). In the following, we use these two terms to quantify the nonlinearity of the system.

We now focus on pump-probe measurements around $\nu = 2/5$ ($B=3.15\,{\rm T}$). We observe, in Fig.~\ref{fig:Fig2}b, that $\mathcal{I}(\omega_{\rm m}, \tau)$ features a fast oscillation modulated by an exponential envelope. This is the expected waveform since $\mathcal{I}(\omega_{\rm m}, \tau)$ is, to lowest order, the autocorrelation signal of the resonance fluorescence emitted by the sample: the carrier frequency of the fast oscillation corresponds to the (undersampled) optical frequency and the characteristic decay time is the polariton coherence time $T_{\rm LP} = 24 \pm 1 \, {\rm ps}$ (dashed black line). The nonlinear contribution $\mathcal{I}(3\omega_{\rm m}, \tau)$, depicted in Fig.~\ref{fig:Fig2}c, also exhibits fast oscillations but its envelope has a more complex structure as a consequence of the interplay between several interfering nonlinear terms, with characteristic decay times $T_{\rm LP}$ and $T_{\rm LP}/3$ that compensate at short delays. Figure~\ref{fig:Fig2}d shows a logarithmic plot of the integrals $\left < \mathcal{I}(\omega_{\rm m}, \tau) \right >_{\tau}  = \int{ \mathcal{I}(\omega_\mathrm{m}, \tau) d\tau} $ and $\left < \mathcal{I}(3 \omega_{\rm m}, \tau) \right >_{\tau} = \int{ \mathcal{I}(3\omega_\mathrm{m}, \tau) d\tau}$ as a function of the average incident power. We observe that the former exhibits a power law with exponent $1.3 \pm 0.3$, which is consistent with the expected linear behavior. By contrast, $\left < \mathcal{I}(3 \omega_{\rm m}, \tau) \right >_{\tau}$ shows a power law with exponent $2.2 \pm 0.3$ that is consistent with the anticipated dependence of third-order nonlinear response, validating that $\left < \mathcal{I}(3 \omega_{\rm m}, \tau) \right >_{\tau}$ is indeed a good measure of the nonlinearity. The observed deviation of the power law exponents from the expected values 1.0 and 2.0 is most likely due to systematic errors on the input power calibration. We emphasize that the measured nonlinearity occurs on timescales that are comparable to the polariton lifetime, which demonstrates that our method allows us to access (fast) polariton-polariton interactions. We also note that the nonlinear response saturates at high optical powers (Fig.~\ref{fig:Fig2}c). The saturation behaviour of $\left < \mathcal{I}(3 \omega_{\rm m}, \tau) \right >_{\tau}$ at high optical powers may be attributed to the saturation of the ${\rm LP}_{\sigma^-}$ red shift induced by a change in $\nu$ (Fig.~\ref{fig:Fig1}b). Saturation could also be a consequence of (slow) light-induced modifications of $n_e$, which may start to play a role at the highest powers investigated \cite{ravets_polaron_2018}.

We further confirm that $\left < \mathcal{I}(3 \omega_{\rm m}, \tau) \right >_{\tau}$ provides a good measure of polariton-polariton interactions, by applying our measurement procedure to a different sample featuring a neutral quantum well ($n_e = 0$). Based on a numerical analysis relying on solving the Gross-Pitaevskii equation (see Supplementary), we quantify the nonlinearity of exciton-polaritons for the neutral quantum well sample. We obtain a value of $13_{-9}^{+18}\,\mathrm{\mu eV}\,\mathrm{\mu m}^2$ for the exciton-exciton interaction strength, which is consistent with values reported elsewhere \cite{munoz-matutano_emergence_2019, delteil_towards_2019, amo_superfluidity_2009, ferrier_interactions_2011, rodriguez_interaction-induced_2016, brichkin_effect_2011, walker_dark_2017}.

\begin{figure}
\centering
\includegraphics[width=120mm]{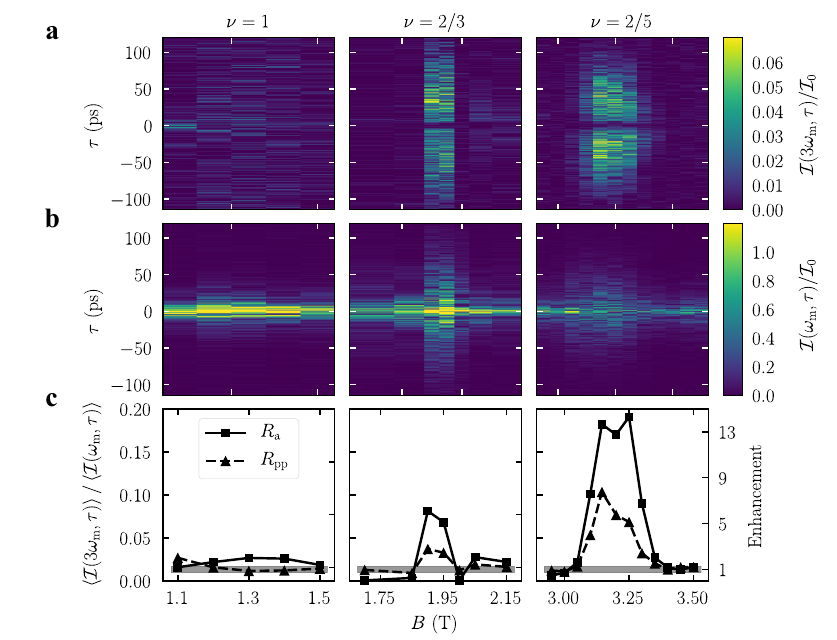}
\caption{{\bf Enhancing interactions between quantum Hall polaritons at fractional filling factors.} {\bf a},~Nonlinear response $\mathcal{I}(3\omega_\mathrm{m}, \tau)/\mathcal{I}_0$ and {\bf b},~linear response $\mathcal{I}(\omega_\mathrm{m}, \tau)/\mathcal{I}_0$ as a function of $B$, in vicinity of filling factors $\nu=1$, $2/3$ and $2/5$. All data is normalized by the same value $\mathcal{I}_0$ as in Fig.~2b. {\bf c},~Enhancement of the nonlinearity, as revealed by the ratio of $3\omega_\mathrm{m}$ and $\omega_\mathrm{m}$ responses. Two different measures of the enhancement strength are plotted, the area $R_\mathrm{a}$ integrated over $\tau$ (squares) and the signal peak-to-peak $R_\mathrm{pp}$ (triangles). In order to obtain a lower bound for the enhancement of interactions, we compare the signal to the noise level. Taking the outermost points in each panel as reference points, the $y$-axis on the right hand side gives the relative enhancement of $U$. The grey shaded area is the standard deviation of the reference points. The excitation power for all measurements is $I_2 = 20\pm 3 $\,nW.}
\label{fig:Fig3}
\end{figure}

Having established that our measurements can be used to reliably determine the polariton-polariton interaction strength, we analyze the evolution of $\left < \mathcal{I}(\omega_{\rm m}, \tau) \right >_{\tau}$ and $\left < \mathcal{I}(3 \omega_{\rm m}, \tau) \right >_{\tau}$ as a function of $\nu$ by measuring $I(t, \tau)$ for different values of $B$. The data consists of three sets centered around filling factors $1$, $2/3$, and $2/5$ that exhibit clear signatures of optical coupling to quantum Hall states, as demonstrated in Fig.~\ref{fig:Fig1}b. For each of the three sets, we tune the cavity to resonance with the singlet attractive polaron channel (polaron content $55 \pm 5 \%$). Then, we tune $B$ to access neighboring filling factors while carefully adjusting the laser frequency to resonantly excite ${\rm LP}_{\sigma^-}$ for every datapoint. The main result of this letter is the remarkable $\nu$-dependence of the nonlinear signal $\mathcal{I}(3\omega_\mathrm{m}, \tau)$ shown in Fig.~\ref{fig:Fig3}a. We observe a strong increase of the nonlinearity at fractional filling factors $\nu = 2/3$ ($B \simeq 1.95\,{\rm T}$) and $\nu = 2/5$ ($B\simeq3.15\,{\rm T}$), as compared to neighboring filling factors. Away from these states, e.g. for $B=3.5\,{\rm T}$, the nonlinearity becomes weaker and eventually comparable to the noise level of our apparatus. This gives clear evidence that polariton-polariton interactions are enhanced around the fractional quantum Hall states $\nu=2/3$ and $\nu=2/5$. In stark contrast, we observe that $\mathcal{I}(3 \omega_{\rm m}, \tau)$ is only marginally modified around the integer filling factor $\nu=1$. Also, another interesting feature is manifest in the delay dependence of $\mathcal{I}(3\omega_{\rm m}, \tau)$ at the fractional filling factors ($\nu = 2/5$ \& $2/3$), which show envelopes with a qualitatively different shape than observed for integer filling factors ($\nu =1$) or expected based on numerical solutions of the driven-dissipative Gross-Pitaevskii equation (see Supplementary section II). This suggests that the nonlinear optical response close to fractional filling factors cannot be captured by a simple Kerr nonlinearity.

We note that, since polaritons interact through their matter part, a change in the polaronic content of the polaritons as a function of $B$ could also lead to a modification of polariton-polariton interactions: polaritons with a higher polaron content will indeed show larger interactions. Since we use a slightly blue-detuned cavity mode, we expect polaron-polaritons to be more polaronic and less photon-like at $\nu=2/5$ due to the reduction of the normal-mode splitting around this filling factor (see Fig.~\ref{fig:Fig1}b). Measurements of the evolution of the polariton effective mass around $\nu=2/5$ have shown that the polaron fraction varies by about a factor of two \cite{ravets_polaron_2018} between $B=3.15~{\rm T}$ and $B=3.5~{\rm T}$. We therefore cannot exclude that part of the enhancement at $\nu=2/5$ stems from the increase in polaron content of the lower polariton branch. However, for $\nu = 2/3$, the polariton splitting (see Fig.~\ref{fig:Fig1}b) and thus the polaron content of the polaritons is only marginally changed and yet we observe a  factor of $\sim 6$ enhancement of the nonlinearity at this filling factor: this indicates that the increase of interaction around fractional filling factors only partially stems from an increase in polaron content of the ${\rm LP}_{\sigma^-}$ mode.

Figure~\ref{fig:Fig3}b shows the evolution of the (mostly linear) term $\mathcal{I}(\omega_{\rm m}, \tau)$ at $\nu = 1$, $2/3$, $2/5$, where we observe another striking feature that coincides with the enhancement of the nonlinearity. For the intermediate pump power used in Fig.~\ref{fig:Fig3}, we find that the characteristic decay time of $\mathcal{I}(\omega_{\rm m}, \tau)$ (i.e. the lower polariton coherence time $T_\mathrm{LP}$) is prolonged at $\nu=2/3$ and $\nu=2/5$. A detailed study of this effect as a function of input power (see Supplementary section III) shows that $T_\mathrm{LP}$ is multiplied by a factor two to three as we increase the input power. The origin of this power dependent enhancement of $T_\mathrm{LP}$ remains unclear.

We summarize our results in Fig.~\ref{fig:Fig3}c, where we provide values of the enhancement of polariton-polariton interactions close to fractional filling factors, obtained by calculating the ratio of the areas $R_{\rm{a}} = \left < \mathcal{I}(3 \omega_{\rm m}, \tau) \right >_{\tau} / \left < \mathcal{I}(\omega_{\rm m}, \tau) \right >_{\tau}$ and the ratio of the signal peak-to-peak values $R_{\rm{pp}}$ 
\footnote{$R_{\rm{pp}} = \left( \max(\mathcal{I}(3\omega_\mathrm{m}, \tau)) - \min(\mathcal{I}(3\omega_\mathrm{m}, \tau))\right) / \left(  \max(\mathcal{I}(\omega_\mathrm{m}, \tau)) - \min(\mathcal{I}(\omega_\mathrm{m}, \tau))\right)$.}.
At $\nu=2/5$, both measures show a significant enhancement of the interaction, of the order of $10^1$, as compared to neighboring filling factors. At a first glance, a possible explanation for the enhancement of the nonlinear response may lie in the filling factor-dependent polariton energy shifts observed in Fig.~\ref{fig:Fig1}b: increasing the incident power changes the local electron density and modifies the polariton Rabi splitting, leading to power-dependent modifications of the polariton energy. However, this explanation is inconsistent with our observation of an increase of the nonlinearity around $\nu=2/3$ despite the fact that the lower polariton energy only changes marginally with $B$ (Fig.~\ref{fig:Fig1}b). Moreover, we do not measure a significant enhancement of the nonlinearity at $\nu =1$, despite strong variations in polariton splitting as we vary $B$ or the pump power.

We also repeated the experiment around the $\nu=1/3$ state ($B = 3.9~\rm{T}$), where we did not observe an enhanced nonlinearity. We speculate that this is due to the suppression of $\sigma^-$ polarized polaron formation at $\nu \leq 1/3$ due perfect spin polarization of the 2DEG, leaving the polariton mode mostly photonic \cite{ravets_polaron_2018}. Moreover, we repeated the experiments with the ${\rm LP}_{\sigma^+}$ resonance at $\nu=2/5$. Remarkably, the nonlinearity was not enhanced. We tentatively explain this observation by the fact that the FQH state is less sensitive to an electron introduced into a higher LL in the case of ${\rm LP}_{\sigma^+}$ resonance, as compared to the lowest LL in the case of ${\rm LP}_{\sigma^-}$ resonance. Last but not least, we measured qualitatively similar results on a second sample with higher electron density ($n_e = 1.4 \times 10^{11}\, \mathrm{cm}^{-2}$) for the (spin-polarized) $\nu = 2/3$ state at $B \simeq 8.5\,{\rm T}$. There, we did observe a comparable enhancement with $\nu$ (see Supplementary section IV).

Strong enhancement of polariton-polariton interactions around FQHE states opens up new perspectives for the study of strongly correlated electron as well as for photonic systems. In light of recent photon correlation measurements in confined exciton-polariton systems \cite{munoz-matutano_emergence_2019,delteil_towards_2019}, the enhancement of interactions by an order of magnitude that we found here shows great potential for reaching the strong polariton blockade regime. Understanding the physical mechanism for enhanced nonlinear response and prolonged polariton coherence times for FQHE states constitutes a very interesting open problem. Our experiments show that despite their qualitatively similar linear optical response, fractional and integer QHE states show strikingly different nonlinear optical signatures: this suggests that nonlinear spectroscopy could reveal signatures of strongly correlated electronic systems that are not accessible by linear optical or transport measurements.

\section*{Methods}
{\bf Sample structure.}
Our sample structure, detailed in \cite{ravets_polaron_2018}, features a 20\,nm modulation doped gallium arsenide (GaAs) quantum well located at the center of a $2\lambda$ $\rm Al_{0.19}Ga_{0.81}As$ microcavity. The two distributed Bragg reflectors consist of 25 (19) pairs of $\rm AlAs / Al_{0.20}Ga_{0.80}As$ layers. The cavity quality factor, measured by white light reflectivity, is $Q \simeq (5.5 \pm 0.1) \times 10^3$. The 2DES shows an electron density $n_e \simeq 0.33 \times 10^{11}\,{\rm cm}^{-2}$ and a mobility $\mu \simeq 1.6 \times 10^6\,{\rm cm}^2{\rm V}^{-1}{\rm s}^{-1}$, as measured by magneto-transport. Note that with these parameters, we conveniently access various integer and fractional quantum Hall states for relatively low magnetic fields $B \leq 5\,{\rm T}$.

{\bf Optical characterization.}
We perform an initial characterization of our sample by polarization-resolved white light reflectivity as a function of $B$. We couple light from a broadband light emitting diode into an optical fiber and shine photons onto the sample placed inside a dilution refrigerator with a base temperature of 30\,mK (see Fig.~\ref{fig:Fig1}a). The light is focused on the sample surface using a low numerical aperture lens (${\rm N.A.=0.15}$) in confocal configuration. Reflected light is then collected by the same fiber and analyzed with a spectrometer. More details on the optical setup can be found in the Supplementary section I.

{\bf Time-resolved measurements.}
One standard method for evaluating interactions between exciton-polaritons in 2D uses a resonant continuous wave excitation laser to monitor the blue-shift experienced by the lower polariton line due to the (Kerr-like) nonlinearity as the polariton population increases. In these experiments, however, one cannot differentiate between the contribution due to fast ($\sim 10\,{\rm ps}$) polariton-polariton interactions, and other unwanted contributions due to the slow ($> 100\,{\rm ps}$) buildup of an excitonic reservoir \cite{stepanov_two-components_2018}. This issue is critical in the context of quantum Hall polaritons since the 2DES electron density is particularly sensitive to optical power due to possible photoionization of DX centers when illuminating the sample: increasing the optical power density may lead to unwanted modifications of $n_e$ and therefore to slow variations of the ($\nu$-dependent) polariton energies, which in turn may prevent us from properly evaluating the interactions. In order to isolate pure polariton-polariton interactions, we use a carefully designed sample structure with reduced sensitivity of $n_e$ to light~\cite{ravets_polaron_2018}, and perform time-resolved experiments in the pulsed-excitation regime in which the pulse duration ($\sim 4\,{\rm ps}$) is shorter than the polariton lifetime ($> 12\,{\rm ps}$). A traditional approach to isolate the nonlinear contribution in four-wave mixing experiments consists in introducing an angle between the two exciting beams in order to generate a background-free nonlinear response at a different angle. However, the requirement for ultra-low temperatures render standard FWM experiments technically challenging to implement in our experimental geometry that uses a fiber coupled scanning confocal microscope in a dilution refrigerator. Instead, we use an electro-optic modulator to sinusoidally modulate the field amplitude $E_1$ in one arm of the interferometer, with angular frequency $\omega_{\rm m} / 2\pi =8\,\mathrm{kHz}$ (see Fig.~\ref{fig:Fig2}a). As a consequence, the power spectral density $\mathcal{I}(\omega,\tau)$ contains terms that oscillate at multiples of the modulation frequency $\omega_{\rm m}$. Expanding the first order terms in Equation (1)

\begin{equation*}
    \left| P^{(1)}_1(t) \sin{(\omega_\mathrm{m}t)} \right|^2 + \left| P^{(1)}_2(t, \tau) \right|^2 + 2\Re{\left( {P^{(1)}_1 (t)}^{*} \sin{(\omega_\mathrm{m}t)}  P^{(1)}_2(t, \tau) \right)},
\end{equation*}
we find that a field autocorrelation term appears at $\omega_\mathrm{m}$. The next order terms are the cross-products between linear and nonlinear polarizations with subscripts denoting fields originating from optical paths 1 and 2

\begin{equation*}
    2\Re{\left(\left( {P^{(1)}_1}^* \sin{(\omega_\mathrm{m}t)}  + {P^{(1)}_2}^* \right)  \left( P_{111}^{(3)}\sin{(\omega_\mathrm{m}t)}^3 + P^{(3)}_{112}\sin{(\omega_\mathrm{m}t)}^2 + P^{(3)}_{122}\sin{(\omega_\mathrm{m}t)} + P^{(3)}_{222} \right)\right)}.
\end{equation*}
In turns out that $3\omega_\mathrm{m}$ is the first frequency for which $P^{(3)}$ contributes to all terms with no background from $P^{(1)}$, so $\mathcal{I}(3\omega_\mathrm{m}, \tau)$ is used to monitor the nonlinear response. Similar techniques have been used for pump probe experiments in collinear geometry \cite{hall_heterodyne_1992, mecozzi_transient_1998, nardin_multidimensional_2013, smallwood_multidimensional_2018}.

{\bf Data availability.} The data that support the plots within this paper and other findings
of this study are available from the corresponding authors upon reasonable request.

\section*{References}
%

\section*{Acknowledgements}
The Authors acknowledge fruitful discussions with Jacqueline Bloch, Antoine Browaeys, Thibault Chervy, Ovidiu Cotlet, Aymeric Delteil, Tobias Grass, Mohammad Hafezi, Emre Togan, Sina Zeytinoglu and Oded Zilberberg. We thank Mirko Lupatini for providing us with a reference sample used for estimating the exciton-exciton interaction in a neutral quantum well. This work was supported by the Swiss National Science Foundation (NCCR Quantum Science and Technology), an ETH Fellowship (S.R.). This project has received funding from the European Research Council under the Grant Agreement No 671000.
\section*{Author contributions}
P.K. and S.R. contributed equally to this work. P.K. and S.R. performed and analyzed the measurements, S.F. and W.W. grew the sample. S.R., M.K. and A.I. supervised the work. P.K., S.R., M.K. and A.I. wrote the manuscript.
\section*{Additional information}
{\bf Supplementary information} for this paper is available online.

{\bf Correspondence and requests for materials} should be addressed to A.I. (imamoglu@phys.ethz.ch) and S.R. (sylvain.ravets@u-psud.fr).

\FloatBarrier

\clearpage
\begin{center}
  \textbf{\large Supplementary material for nonlinear optics in the fractional quantum Hall regime}\\
\end{center}
\setcounter{equation}{0}
\setcounter{figure}{0}
\setcounter{table}{0}
\setcounter{page}{1}
\setcounter{footnote}{0}

We provide, in this supplementary material, a more complete description of the experimental setup and data analysis procedure. We first detail the experimental techniques used for sample characterization and for measuring polariton-polariton interactions. We then describe the reference measurement used to benchmark our experimental setup. Based on a simple model, we estimate the exciton-exciton interaction strength in a neutral quantum well to amount to $13_{-9}^{+18}\,\mathrm{\mu eV}\,\mathrm{\mu m}^2$, consistent with other reports \cite{ciuti_role_1998, tassone_exciton-exciton_1999, munoz-matutano_emergence_2019, delteil_towards_2019, amo_superfluidity_2009, ferrier_interactions_2011, brichkin_effect_2011,  rodriguez_interaction-induced_2016, walker_dark_2017}. We finally present a detailed study of the evolution of the polariton coherence time around fractional filling factors.
\FloatBarrier

\section{Experimental setup}
\subsection{White light reflectivity}
\begin{figure}
\centering
\includegraphics[width=115mm]{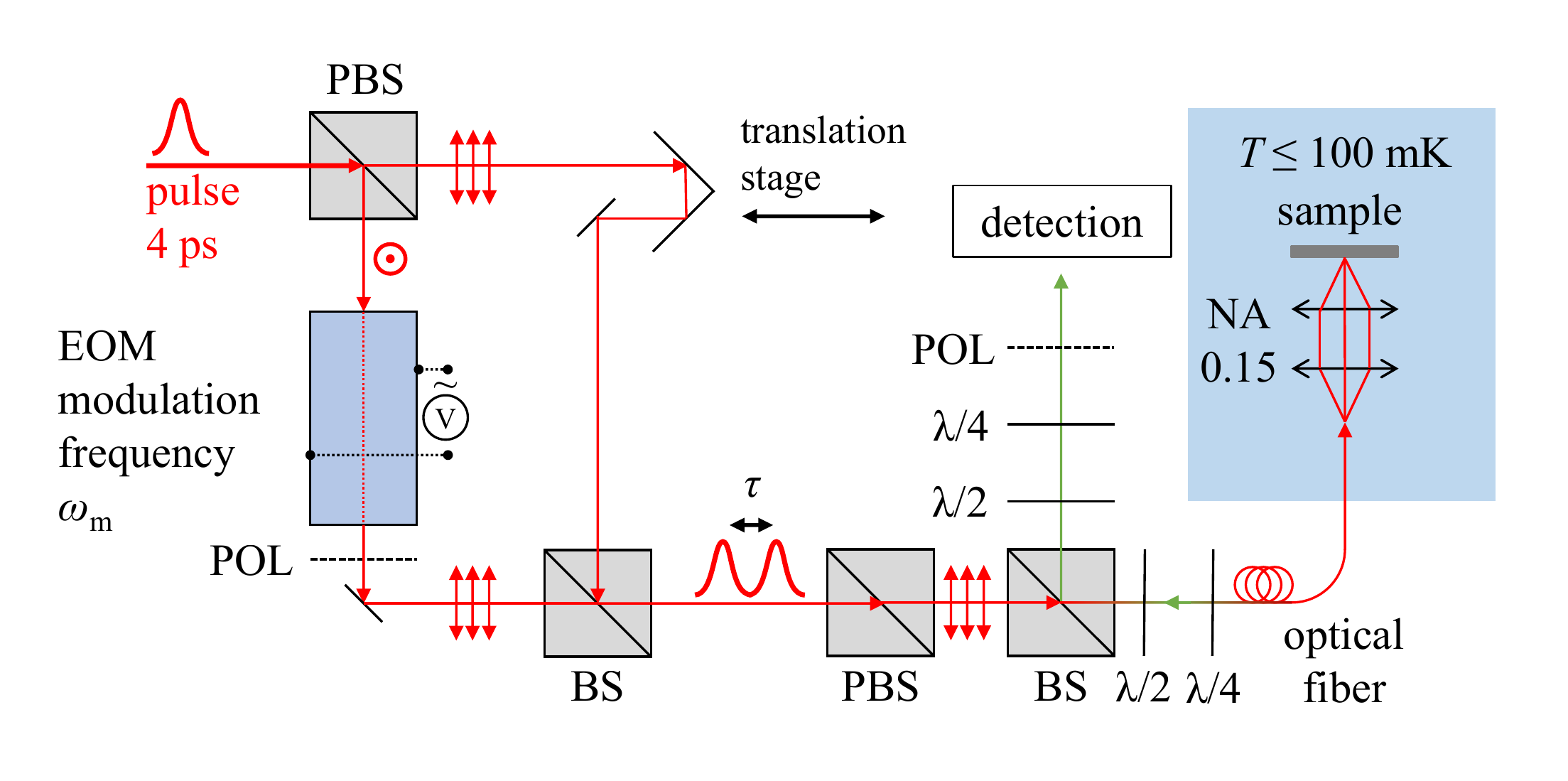}
\caption{{\bf Experimental setup.} Schematic of the interferometer used for measuring the nonlinear response of the system.
}
\label{fig:FigS1}
\end{figure}

For sample characterization, we first record polarization-resolved white light reflectivity spectra. The main goal here is to find the resonance between the singlet-polaron channel and the cavity mode, and also to adjust polarization optics both in excitation and in detection. We shine on the order of a few nanowatts of optical power from a broadband light emitting diode\footnote{Exalos SLED centered at 820\,nm.} onto the sample. For that measurement, we block one of the arms of the interferometer shown in Fig.~\ref{fig:FigS1}. The light is then coupled to a single mode fiber routed to the sample placed inside a dilution refrigerator\footnote{Leiden Cryogenics MCK 76-400 Attocube.} with a base temperature of 30\,mK. Monitoring the polariton spectrum around $\nu=2/5$ while increasing the temperature suggests that the electron temperature is lower than $200\,{\rm mK}$. The light is focused onto the sample surface using a low numerical aperture lens (${\rm N.A.=0.15}$). Reflected light is then collected by the same fiber and analyzed in the detection path (green line in Fig.~\ref{fig:FigS1}). The collected light is finally sent to a spectrometer\footnote{Princeton Instruments SP2500i.} equipped with a nitrogen-cooled CCD\footnote{Princeton Instruments liquid nitrogen cooled CCD.}. Spectra $s$ are recorded for different values of $B$. By combining the resonance-free spectral regions obtained for different values of $B$ we construct a background spectrum $r$. We then calculate $1 - s/r$, to obtain the background corrected spectra shown in Fig.~1 of the main text. Importantly, we are also able, based on the resonances observed in the white light reflectivity spectrum, to optimize carefully the polarization of the input light field to be either right-hand circularly polarized $(\sigma^+)$ or left-hand circularly polarized $(\sigma^-)$.

\subsection{Time resolved pump-probe measurements}
For measuring the interactions, we perform time-resolved spectroscopic measurements in a pump-probe configuration. We split a few picosecond laser pulse\footnote{Coherent MIRA 900 mode locked Ti:Sapphire laser.} into two paths, and introduce a variable delay $\tau$ before recombining them onto a beam splitter (see Fig.~\ref{fig:FigS1}). One of the two arms of this interferometer is equipped with an electro-optic modulator (EOM)\footnote{Linos LM0202.} used for modulating the pump amplitude. Note that access to the sample is restricted to reflection in colinear configuration: to separate the resonance fluorescence emitted by the sample from laser light that reflects off the surface, we use linearly polarized light in excitation and cross-polarized detection. In this way, we suppress the background laser light by 3-4 orders of magnitude, leaving the resonance fluorescence as the dominant contribution to the detected signal.

To distinguish the linear response from the (weaker) nonlinear response, we use an EOM placed between two crossed polarizers as an amplitude modulator. We apply a saw-tooth voltage profile to the EOM to create an electric field amplitude with sine modulation at frequency of $\omega_\text{m} / 2\pi =  8011 \text{\,Hz}$. We optimize the EOM input voltage profile and the EOM alignment to realize a clean sine modulation at this frequency, with less than a percent of higher order harmonic contributions at $2, 3, 4 \times \omega_\text{m}$. We finally couple the reflected signal to a single-mode fiber and send it to an APD\footnote{Excelitas Photon Counting Module SPCM-AQRH-14-FC.}, making sure that the count rate is well in the linear regime of the APD ($\sim 80000~\mathrm{s}^{-1}$). 

The measurement procedure goes as follows. For a chosen magnetic field, we first adjust the cavity energy to {55 $\pm$ 5\%} exciton-polaron content for $\text{LP}_{\sigma^-}$. We then tune the laser pulse central energy to the $\text{LP}_{\sigma^-}$ resonance and suppress the reflected laser light. We note that when scanning $B$ (typically by few $100\,{\rm mT}$) around a given filling factor (e.g. $\nu=2/5$) the singlet polaron resonance energy shift is small compared to the polariton normal mode splitting\footnote{This does not imply a small energy shift for the polariton formed from this singlet polaron, as presented in Fig.~1 of the main text.}: as a consequence, we can keep the cavity energy constant while studying a given filling factor. We also keep the average intensities of pump $(1)$ and probe $(2)$ equal, which was found to result in a good signal to noise ratio. For each time delay $\tau$, we acquire photon counts for $1\,\mathrm{s}$, with the exception of Fig.~2 of the main text, where we used $10\,\mathrm{s}$ acquisition time. The APD sample frequency is $1\,\mathrm{MHz}$, but data binning then leads to an effective sampling frequency of $9 \omega_m$. We then calculate the absolute value of the Fourier transform $I(t, \tau) \mapsto \mathcal{I}(\omega, \tau)$ for the recorded time traces, from which we extract frequency bins corresponding to the first multiples of $\omega_\mathrm{m}$. After background removal, we finally obtain $\mathcal{I}(\omega_\mathrm{m}, \tau)$ and $\mathcal{I}(3\omega_\mathrm{m}, \tau)$ (where the background is derived by averaging $\mathcal{I}(\omega,\tau)$ in vicinity of the frequency of interest).

We perform a test experiment by red detuning the laser so our sample acts as a simple mirror. We adjust the detection polarizers such that the ADP count rate matches the one used in the main experiment. By applying the same experimental procedure to the signal, we observe that $\mathcal{I}(\omega_\mathrm{m}, \tau)$ corresponds to the laser pulse spectrum, whereas no signal is observed at the frequency $3\omega_\mathrm{m}$. This excludes the possibility that the detector or any other optical elements in the setup contribute to the observed nonlinear signal. In another test experiment, we check the behaviour of $\mathcal{I}(3\omega_\mathrm{m}, \tau)$ in response to cavity-polaron detuning. We observe that the nonlinear signal decreases when we red-detune the cavity with respect to the polaron energy: this is the expected behavior since the polaron content of the polaritons is decreased, and the polaritons are thus more photon-like.

\section{Measurement and model of exciton-polariton interactions}
We present, in this section, measurements on an undoped sample ($n_e=0$), which does not contain a 2DEG but only a neutral (intrinsic) QW. The goal is to measure the exciton-exciton interaction strength in a standard single QW (thickness $15\,\mathrm{nm}$) embedded in a DBR cavity and to compare it with known values of the interaction in order to establish our measurement technique as a viable tool for measuring interactions. In order to be able to have a single spin species and comparable conditions with the experimental work presented in the main text, namely linearly polarized excitation, orthogonal detection and $\sigma^{-}$ polarized resonance, we performed this measurement under a $10\,\mathrm{T}$ magnetic field \footnote{For $B=10~{\rm T}$, we do not expect the magnetic field to strongly influence the strength of exciton-exciton interactions since the exciton Bohr radius is still of the same order as the magnetic length.}. To quantify the interaction, we compare our measurements with solutions of a single mode Gross-Pitaevskii equation \cite{wouters_excitations_2007, keeling_spontaneous_2008}, which requires that we estimate the polariton number $N=|\psi(t)|^2$ created by laser excitation in our experiment. Since we use pulsed resonant excitation with a low excitation duty cycle, we do not include in our model contributions from a dark exciton reservoir.

\subsection{Setup calibration}
\begin{figure}
\centering
\includegraphics[width=5in]{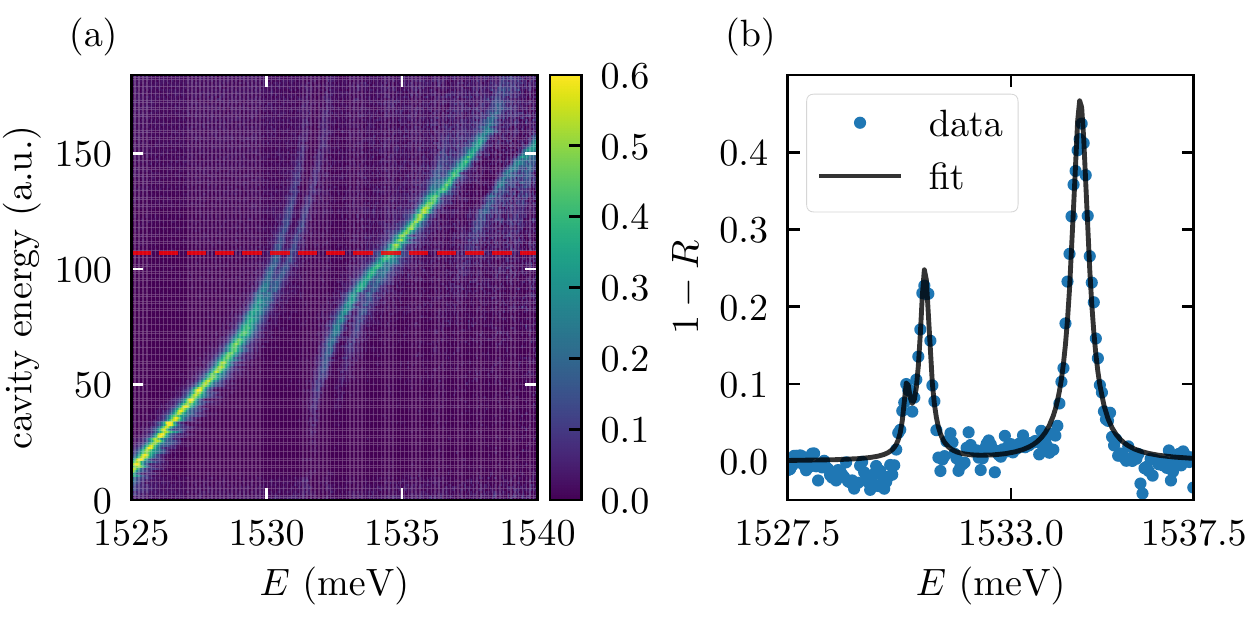}
\caption{{\bf White light reflectivity measurements.} (a) Evolution of the reflectivity spectra as we tune the cavity energy across the exciton resonance. The red line marks the cavity energy for the spectrum shown in the right panel. (b) Background subtracted spectrum (blue dots). The black line shows Lorentzian fits to the spectrum. From the peak areas, we determine the exciton content $|X|^2=0.7$. The lower polariton amplitude is $\eta_c=0.24$.
}
\label{fig:FigS2}
\end{figure}

We estimate, in this section, the polariton occupation number $N$ under picosecond laser pulsed excitation. The (Gaussian) laser pulse has a measured FWHM that is equal to $460\,\mathrm{ueV}$, while the LP resonance (Lorentzian) showed a FWHM of $230\,\mathrm{ueV}$. From this, we estimate the spectral overlap between the laser pulse and the LP line $\eta_{s} = 0.5$. Based on white light reflectivity data (see Fig.~\ref{fig:FigS2}), we also estimate the exciton content $|X|^2=0.7$ and the coupling efficiency into the LP mode $\eta_c=0.24$ (see Lorentzian fits in Fig.~\ref{fig:FigS2}). Knowing the laser power impinging on the sample surface, we can estimate $$N = \eta_s \eta_c n_\mathrm{ph} \, ,$$ where $n_\mathrm{ph} = {p_\mathrm{cw} }/{\left( \hbar \omega_\mathrm{L} f_\mathrm{rep} \right)}$ is the photon number per pulse, $p_\mathrm{cw}$ is the average input power, $f_\mathrm{rep} = 76\,\mathrm{MHz}$ is the pulse repetition rate and $\omega_\mathrm{L}$ is the laser center frequency. 

\subsection{Fit of exciton-polariton interaction strength}
To model the experiment described in section I.B, we use a single mode Gross-Pitaevskii equation for the lower polariton wavefunction: 
\begin{equation} \frac{d\psi(t)}{dt} = - \frac{\gamma}{2} \psi(t) - ig|\psi(t)|^2\psi(t) + F(t,\tau,t_{\rm mod}) \, , \end{equation}
where $g$ is the nonlinearity and $\gamma=0.1\,\mathrm{ps^{-1}}$ is the cavity decay rate. The (modulated) drive term reads:
\begin{equation*} F(t, \tau, t_{\rm mod})=A_1(t_{\rm mod})G(t) + A_2G(t-\tau) e^{i\omega_\mathrm{L} \tau} \, , \end{equation*}
where $G(t)$ and $G(t-\tau)$ are $4\,\mathrm{ps}$ FWHM Gaussian envelopes delayed by $\tau$, $A_1$ is the (modulated) amplitude of the first pulse and $A_2$ the (constant) amplitude of the delayed pulse. In the simulation, we adjust the pulse amplitudes $A_1$ and $A_2$ to match the intracavity polariton number $N$ we estimated in the previous section. The pulse intensities, averaged over a modulation cycle, are chosen to be equal.

\begin{figure}
\centering
\includegraphics[width=6in]{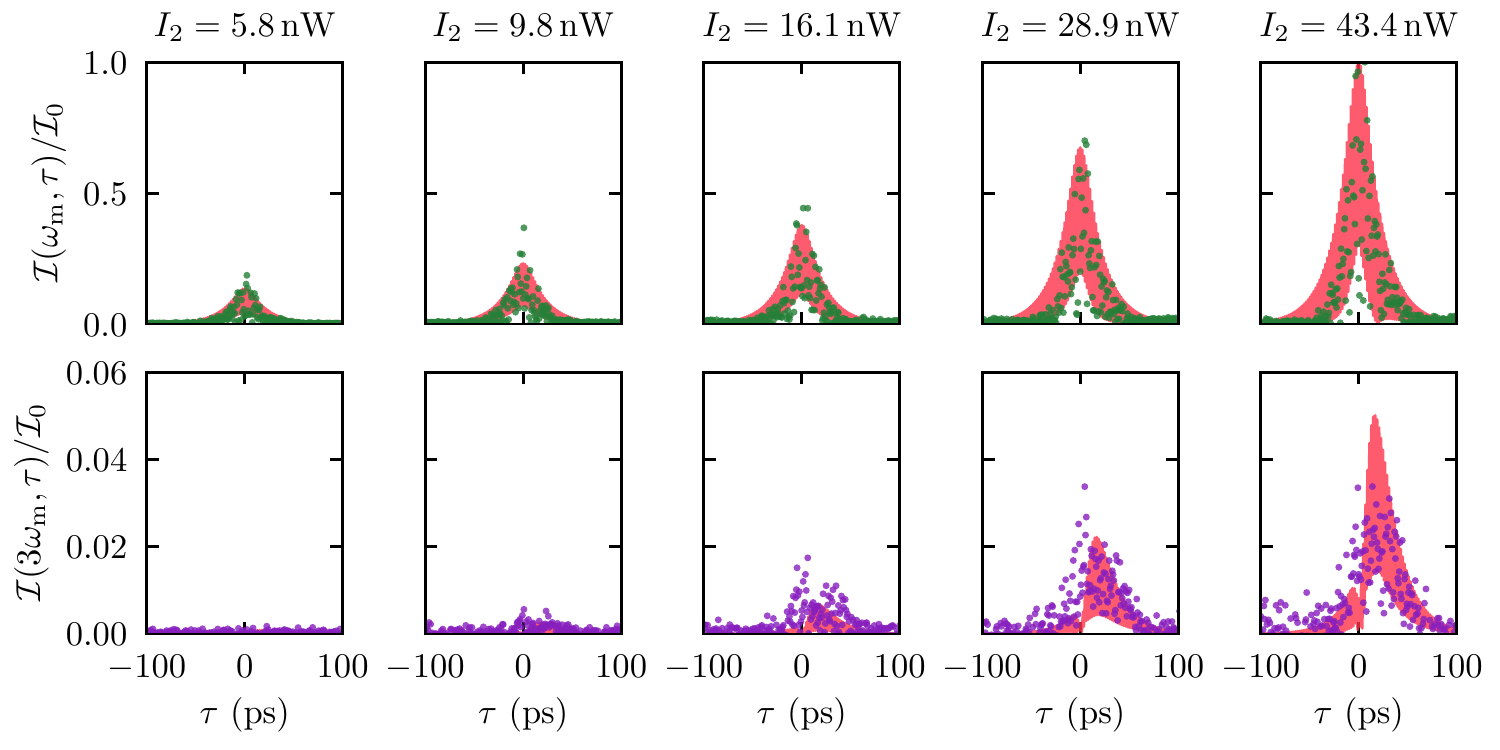}
\caption{{\bf Comparing data from undoped QW sample to GPE.} Top row: comparison between the measured (green circles) and calculated (red shaded area) $\mathcal{I}(\omega_\mathrm{m}, \tau)$ for different input powers, used to calibrate the detection efficiency $\phi$. Bottom row: comparison between the measured (purple circles) and calculated (red shaded area) $\mathcal{I}(3 \omega_\mathrm{m}, \tau)$ for different input powers, yields a value of $g=0.54\,\mathrm{\mu eV}$ for the polariton interaction strength.}
\label{fig:FigS3}
\end{figure}

We calculate $\psi$ for every $\tau$ and we repeat this procedure for different values of $A_1(t_{\rm mod})=\sqrt{I_1}  \sin(\omega_\mathrm{m} t_{\rm mod})$, thus simulating the experimental procedure described in section I.B. We then Fourier transform $\psi$ to obtain the calculated Fourier spectrum $\mathcal{I}_{\mathrm{model}}(\omega, \tau)$ that we directly compare to the experiment as shown in Fig.~\ref{fig:FigS3}. In the end, the simulation includes only two free parameters: the interaction strength $g$ and a global scaling factor $\phi$ that accounts for the finite detection efficiency in our experiment ($\mathcal{I}(\omega, \tau) = \phi \mathcal{I}_{\mathrm{model}}(\omega, \tau)$ where $\phi$ is common to all values of $\omega$ and $\tau$). We determine the parameter $\phi$ by fitting $\mathcal{I}_{\rm model}(\omega_\mathrm{m}, \tau)$ to our experiments. Then, we obtain $g$ by adjusting $\mathcal{I}_{\rm model}(3\omega_m, \tau)$ to best reproduce our measurements. Note that, given the estimate of $N = |\psi|^2 \propto \mathcal{I}(\omega_\mathrm{m}, \tau)$, the information about $g$ is contained in the ratio of $\mathcal{I}(3\omega_\mathrm{m}, \tau) / \mathcal{I}(1\omega_\mathrm{m}, \tau)$, where $\phi$ drops out.

We show, in Fig.~\ref{fig:FigS3}, the results of our fit, which yields a value of $g=0.54\,\mathrm{\mu eV}$ for the polariton interaction strength. To convert this single mode interaction energy into a 2D polariton-polariton interaction constant $U$, we multiply $g$ by the polariton mode area $A$: $U=A \times g$. Based on the numerical aperture of our objective ${\rm N.A.=0.15}$, we expect the excitation beam to extend over $A =11\,\mathrm{\mu m}^2$, which results in $U=6.2\,\mathrm{\mu eV \mu m^2}$. Finally, we estimate the exciton-exciton interaction strength by dividing $U$ by the exciton content squared:  \begin{equation} U_X=A\cdot g/|X|^4 = 13^{+18}_{-9}\,\mathrm{\mu eV \mu m^2}. \end{equation} This result is compatible with other values reported in the literature \cite{ciuti_role_1998, tassone_exciton-exciton_1999, munoz-matutano_emergence_2019, delteil_towards_2019, amo_superfluidity_2009, ferrier_interactions_2011, brichkin_effect_2011,  rodriguez_interaction-induced_2016, walker_dark_2017}.

The largest sources of errors on the measurement of $U_X$ originate from the estimate of $N$ and $A$. Our estimate of $N$ could easily be off by a factor of two. Additionally, our estimate of $A$ might deviate from the spot size estimate due to polariton diffusion and will eventually be modified by the exciton-exciton interactions. Altogether, this leads to the error estimate\footnote{Note that $U_X$ is inversely proportional to the estimated polariton number $N$.} of Eq.~2. We observe, in Fig.~\ref{fig:FigS3}, a small deviation in the power dependence between experiment and fit. This discrepancy is due to a systematic calibration error of the input power (which also led to the observed deviation in the slopes measured in Fig.~2(d) of the main text). However, the resulting systematic error on $g$ is small compared to the first two contributions. We expect our measuring technique to give more accurate results in experimental geometries that allow imaging the polariton cloud in real-space and measuring in transmission.

We finally emphasize that, in Fig.~3 of the main text, we quantified the enhancement of interactions by comparing the nonlinear response of the system at fractional filling factors to the nonlinear response of the system at generic filling factors. Having a baseline that corresponds to the neutral quantum well case would require to evaluate how exciton-exciton interactions are affected by the 2DES at generic filling factors. This will be the scope of future (theoretical and experimental) work, that will aim at clarifying the role of the polaron quasiparticle weight, the electron screening and also the effect that fractional quasiparticles have on the strength and range of the interactions.

\section{Increase in polariton coherence time}
\begin{figure}
\centering
\includegraphics[width=6in]{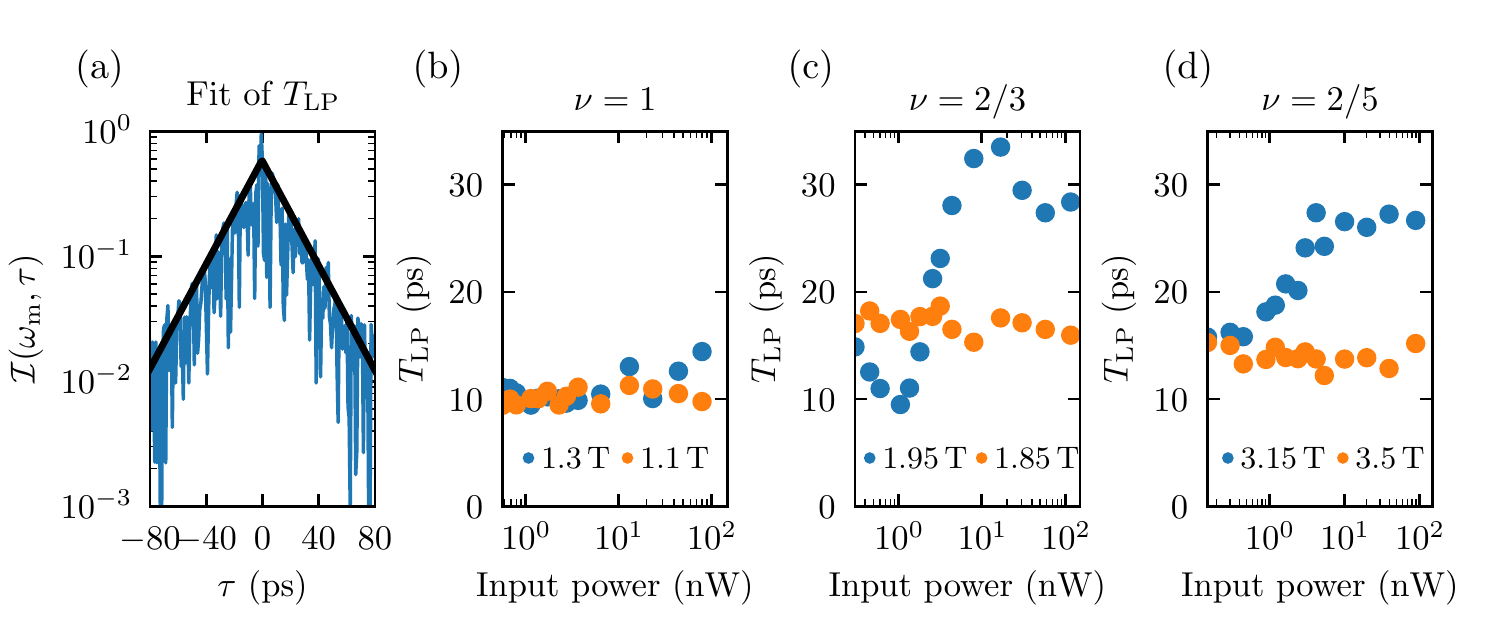}
\caption{{\bf Increase in polariton coherence time with input power at fractional quantum Hall states.} (a) Extraction of $T_\text{LP}$, showing an exemplary linear response in a logarithmic plot with the fit to the envelope in green. The inverse slope corresponds to $T_\text{LP}$. (b-d) Dependence of $T_\text{LP}$ on input power for the filling factors considered in the main text. Blue circles correspond to the magnetic field at the quantum Hall state, orange circles to a magnetic field nearby.
}
\label{fig:FigS4}
\end{figure}

We have observed, in Fig.~3b of the main text, an interesting evolution of the (mostly linear) term $\mathcal{I}(\omega_\mathrm{m}, \tau)$ as we tune $B$: the characteristic decay time of $\mathcal{I}(\omega_\mathrm{m}, \tau)$ (i.e. the polariton coherence time $T_{\rm LP}$) increases for $\nu=2/3$ and $\nu=2/5$. We present, in this section, a detailed study of this effect versus input pump power. We show, in Fig.~\ref{fig:FigS4}(a), a semilog plot of a typical trace $\mathcal{I}(\omega_\mathrm{m}, \tau)$ (blue). To extract $T_\mathrm{LP}$, we fit the envelope of this trace by a double exponential decay (black line): $T_\mathrm{LP}$ is directly given by the exponential decay time. In Fig.~\ref{fig:FigS4}(b)-(d), we plot the fitted values of $T_{\rm LP}$ as a function of input pump power around filling factors $\nu=1, \, 2/3, \, 2/5$. In every panel, we show a dataset recorded when $B$ is tuned to the corresponding quantum Hall states (blue points), and another dataset recorded at a slightly different filling factor (orange points).
At exactly $\nu=2/3$ and $\nu=2/5$, we observe that $T_{\rm LP}$ first increases sharply and then stabilizes at a value two to three times larger to its low-power value. This increase of $T_{\rm LP}$ coincides with the enhancement (and high power saturation) of the nonlinearity at filling factors $\nu=2/3$ and $\nu=2/5$ discussed in the main text. In stark contrast, slightly away from these filling factors, as well as for $\nu=1$, $T_{\rm LP}$ stays relatively stable around its low power value. In another set of experiments, we extracted $T_{\rm LP}$ by measuring the Lorentzian width of $LP_{\sigma -}$ in white light reflectivity spectra as a function of input power; this study (not shown here) led to the same observations. These results suggest a nonlinear behavior of $\mathcal{I}(\omega_\mathrm{m}, \tau)$ at fractional filling factors. However, monitoring the average value $\left < \mathcal{I}(\omega_{\rm m}, \tau) \right >_{\tau}$ versus power (see Fig.2d of the main text) shows that $\left < \mathcal{I}(\omega_{\rm m}, \tau) \right >_{\tau}$ remains linear in excitation power. At this stage, the origin of this power dependent enhancement of $T_\mathrm{LP}$ thus remains unknown. While the measured increase in nonlinearity is clearly an advantage for implementing strongly interacting polaritons, it is unclear whether the observed (high power) increased coherence time (and thus decreased linewidth) could also be beneficial for realizing polariton blockade.

\section{Repeating the measurements for a higher density sample}
\begin{figure}
\centering
\includegraphics[width=6in]{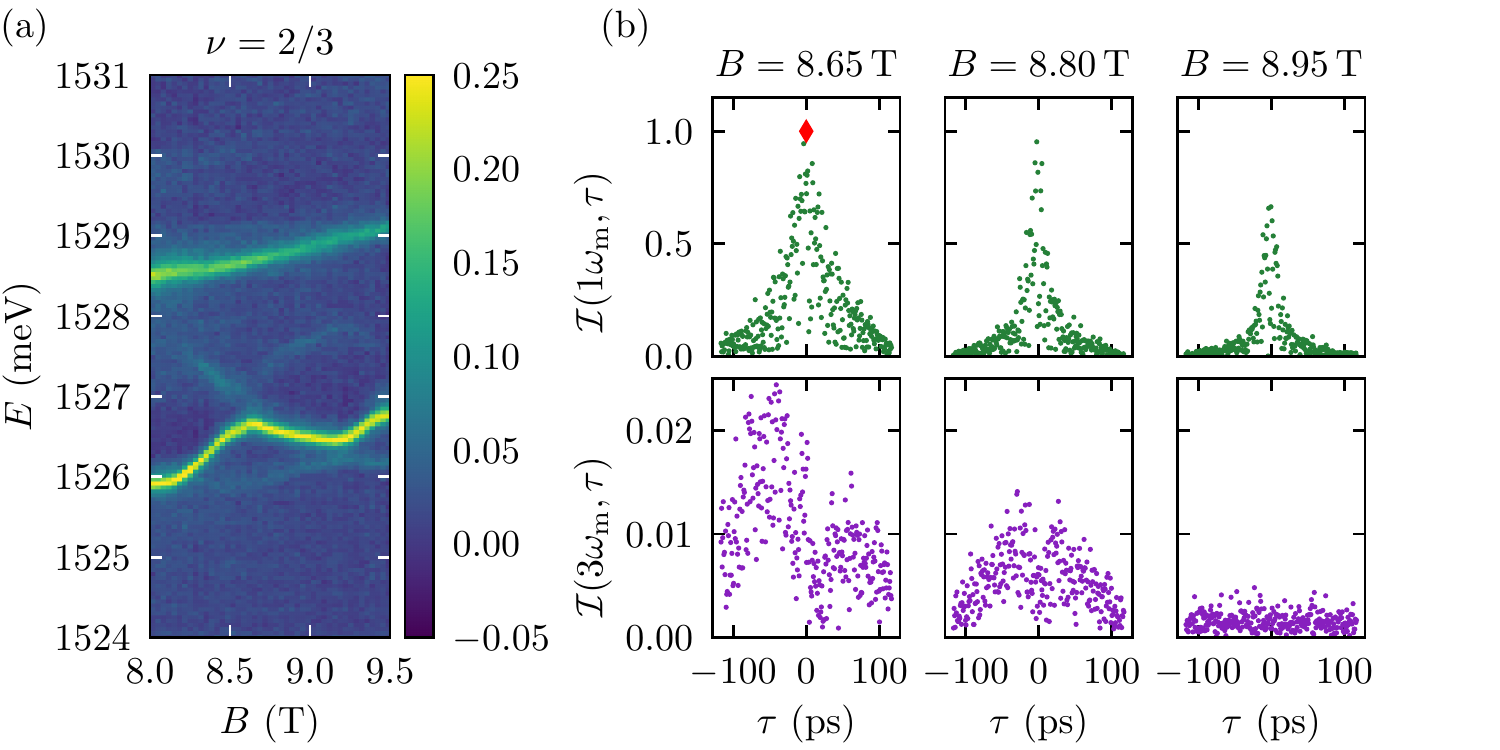}
\caption{{\bf Additional data from another sample.} (a) White light reflectivity spectrum recorded using $\sigma^{-}$ polarized light. At $B=8.6\,$T, the optical signature of $\nu=2/3$ shows as a reduction of the polariton splitting around 1527\,meV (note that the upper polariton is particularly faint). (b) Pump-probe experiment around filling factor $\nu=2/3$. The top row shows $\mathcal{I}(\omega_\mathrm{m}, \tau)$ while the bottom row shows $\mathcal{I}(3\omega_\mathrm{m}, \tau)$. All data has been normalized to the maximal value of $\mathcal{I}(\omega_\mathrm{m}, \tau)$ at $B=8.65\,$T (red diamond). The integration time was chosen equal to 10\,s and the input power was $35 \pm 5 \,$nW.
}
\label{fig:FigS5}
\end{figure}

We repeat our measurement using another sample with higher density electron density of the 2DES ($n_e = 1.4 \times 10^{11}\, \mathrm{cm}^{-2}$). We plot, in Fig.~\ref{fig:FigS5}(a), the white light reflectivity measurement recorded in $\sigma^-$ polarization around filling factor $\nu=2/3$ ($B \simeq 8.6\,$T). Note that, contrary to the lower electron density sample presented in Fig.~1b of the main text, the $\nu=2/3$ state is spin-polarized at this magnetic field. This is observed in the absorption spectrum of Fig.~\ref{fig:FigS5}(a), that resembles the spectrum recorded for the (spin-polarized) state at $\nu=2/5$ in the low density sample sample (see the sharp reduction of normal mode splitting at $B=8.65 \, {\rm T}$). We note however that the coupling efficiency of incident light into the polariton modes was reduced in the high-density sample. We show, in Fig.~\ref{fig:FigS5}(b), the results of our time-resolved four-wave mixing measurement around $\nu = 2/3$. We observe a clear nonlinear response $\mathcal{I}(3\omega_\mathrm{m}, \tau)$ when $\nu$ is tuned to 2/3 exactly. As we go away from $\nu=2/3$, the nonlinearity decreases (bottom row). This behavior is very similar to the one presented in the main text for the low-density sample, since we observe a strong dependence on the filling factor of the nonlinear response $\mathcal{I}(3\omega_\mathrm{m}, \tau)$. The top row also shows the linear response $\mathcal{I}(\omega_\mathrm{m}, \tau)$ for comparison, where we observe the increase of $T_{\rm LP}$ for $\nu=2/3$ (see variations in the exponential decay time of $\mathcal{I}(\omega_\mathrm{m}, \tau)$). With this measurement, we demonstrate the repeatability of our measurement, using another sample with higher electron density. A quantitative comparison of the interaction strengths between the two samples, is however rendered difficult due to the different experimental conditions met with the two samples, and in particular due to the strong difference in coupling efficiency of incident light into the polariton modes in the two samples.

\end{document}